\documentclass[conference]{IEEEtran}
\pdfoutput=1
\usepackage{subcaption}
\usepackage{tikz}
\usepackage{nicefrac}
\usepackage{csquotes}
\usepackage{mathtools,amssymb,amsthm}
\usepackage{csvsimple}
\usepackage[noadjust]{cite}
\usepackage{enumitem}
\usepackage{siunitx}
\usepackage[bookmarks=false,hidelinks]{hyperref}

\hypersetup{
	pdftitle={How to Efficiently Handle Complex Values? Implementing Decision Diagrams for Quantum Computing},
	pdfauthor={Alwin Zulehner, Stefan Hillmich, and Robert Wille}
}

\usetikzlibrary{backgrounds,fit,decorations.pathreplacing,calc,matrix}
\newtheorem{example}{Example}

\newcommand{\ket}[1]{\ensuremath{\left|#1\right\rangle}}

\usepackage[keeplastbox]{flushend}

\begin{document}
\title{How to Efficiently Handle Complex Values?\\{\huge Implementing Decision Diagrams for Quantum Computing}}

\date{}
\author{\IEEEauthorblockN{Alwin Zulehner, Stefan Hillmich, and Robert Wille}
	\IEEEauthorblockA{Institute for Integrated Circuits, Johannes Kepler University Linz, Austria \\
alwin.zulehner@jku.at \hspace{0.45cm} stefan.hillmich@jku.at \hspace{0.45cm} robert.wille@jku.at
}}

\maketitle

\begin{abstract}
Quantum computing promises substantial speedups by exploiting quantum mechanical phenomena such as superposition and entanglement. Corresponding design methods require efficient means of representation and manipulation of quantum functionality. 
In the classical domain, decision diagrams have been successfully employed as a powerful alternative to straightforward means such as truth tables. 
This motivated extensive research on whether decision diagrams provide similar potential in the quantum domain---resulting in new types of decision diagrams capable of substantially reducing the complexity of representing quantum states and functionality.
From an implementation perspective, many concepts and techniques from the classical domain can be re-used in order to implement decision diagrams packages for the quantum realm. However, new problems---namely how to efficiently handle complex numbers---arise. 
In this work, we propose a solution to overcome these problems. 
Experimental evaluations confirm that this yields improvements of orders of magnitude in the runtime needed to create and to utilize these decision diagrams. 
The resulting implementation is publicly available as a quantum DD package at \url{http://iic.jku.at/eda/research/quantum_dd}. 
\end{abstract}

\section{Introduction}
\label{sec:introduction}

Quantum computing~\cite{NC:2000} promises significant speedups for certain problems, e.g.,~integer factorization \cite{DBLP:journals/siamcomp/Shor97}, database search \cite{DBLP:conf/stoc/Grover96}, and quantum chemistry~\cite{lanyon2010towards}. 
Unlike bits in a classical computer, which can only assume one of the two basis states \(0\) and \(1\), \emph{qubits} in a quantum computer can be in an (almost) arbitrary superposition of both. 
Superposition in combination with other quantum phenomena such as entanglement and phase shifts allows for an exponential speedup compared to classical computers in the best case. Not surprisingly, this triggered a huge interest in the utilization of quantum devices which, recently, led to first realizations provided by large commercial players such as IBM and Google. 

These developments are currently progressing to a point where straightforward automated approaches for, e.g.,~synthesis, simulation, or verification, become indispensable
since
corresponding quantum states as well as quantum operations are mathematically described through state vectors and unitary matrices which grow exponentially in size with respect to the number of involved qubits.

In the classical domain, the design automation community successfully addressed such challenges by introducing decision diagrams which, in many cases, allow for a compact representation of functionality. Impressive accomplishments in the '90s, e.g.,~with \emph{Binary Decision Diagrams} (BDDs,~\cite{Bry:85}), \emph{Binary Moment Diagrams} (BMDs,~\cite{bryant1995verification}), or \emph{\mbox{Zero-suppressed} Decision Diagrams} (ZDDs,~\cite{Min:93}), are examples of the potential of those representations. Corresponding implementations (usually denoted \emph{DD packages}) as provided by Fabio Somenzi's~CUDD package~\cite{somenzi2015cuddpackage}, the \mbox{Word-level-DD} package~\cite{herbstritt2004wld}, or Donald Knuth's~BDD package~\cite{knuth2009art} affect the development of design tools and methods until today.
Motivated by that, in the past years, researchers spent considerable efforts in the investigation of whether decision diagrams can also be utilized in the quantum computing realm. This led to several theoretical and mathematical concepts of decision diagrams such as \emph{\mbox{X-decomposition} Quantum Decision Diagrams}~(XQDDs,~\cite{WLTK:2008}), \emph{Quantum Decision Diagrams} (QDDs,~\cite{abdollahi2006analysis}), \emph{Quantum Information Decision Diagrams} (QuIDDs,~\cite{VRMH:2003}), or \emph{Quantum Multiple-valued Decision Diagrams} (QMDDs,~\cite{MT:2006,DBLP:journals/tcad/NiemannWMTD16})---leading to more efficient methods for the design tasks outlined above, i.e., 
synthesis~\cite{niemann2016logic, zulehner2017one,niemann2018cliffordt},
simulation~\cite{viamontes2004high,zulehner2017advanced,zulehner2019matrix}, or
verification~\cite{WLTK:2008,yamashita2008ddmf,niemann2014equivalence}.

However, besides the mathematical concepts, efficient \emph{implementations} of the corresponding decision diagrams for quantum computing are needed in order to eventually utilize them for design automation at large scale---an issue that has not explicitly been addressed thus far. 
Key concepts required for implementing DD packages---such as unique tables, garbage collection with reference counts, or compute tables---are already known from the classical domain (which, also in the '90s, have explicitly been investigated, e.g.,~in~\cite{Bry:86,BRB:90,SB:96}).
While these implementation techniques can be directly incorporated into decision diagrams addressing quantum computing, the quantum realm additionally requires an efficient handling of complex numbers. These complex numbers introduce several new problems such as how to keep numerical stability, how to efficiently store nodes in unique tables, as well as how to store reoccurring operations in compute tables. None of these issues have explicitly been considered thus far.

In this work, we provide details on how to efficiently implement a DD package for quantum computing addressing these problems.
Established concepts known from decision diagrams in the classical domain are re-used where applicable, while new implementation techniques for handling complex values efficiently 
are described in detail.
Experimental results confirm that these efficient implementation techniques yield improvements of orders of magnitude with respect to runtime
compared to the best known implementations available today. An implementation of the resulting DD package is publicly available at \url{http://iic.jku.at/eda/research/quantum_dd}.

\vspace{200cm}

The remainder of this paper is structured as follows: 
Section~\ref{sec:background} briefly reviews the basics of quantum computing and applicable decision diagrams.
In Section~\ref{sec:implementation}, we recapitulate implementation techniques used for classical DD packages before discussing problems related to the extension of them into the quantum domain.
Based on that, Section~\ref{sec:handling-complex-weights} describes in detail how the corresponding problems can efficiently be  handled---yielding an efficient implementation of a DD package for quantum computing.
Eventually, the contributions of this work have been incorporated into an implementation of a DD package for quantum computing whose performance is evaluated and compared to the state of the art in Section~\ref{sec:results}. Finally, the paper is concluded in Section~\ref{sec:conclusions}.

\section{Background}
\label{sec:background}

In this section, we review the basics of quantum computing and decision diagrams for representing quantum functionality.

\subsection{Quantum Computing}
\label{sec:quantum_comp}

Computations in the quantum realm use qubits, which can assume more states than the basis states (here, written as \ket0 and \ket1 using Dirac-notation) used in classical computations.
A quantum state \(\ket{\psi}\) is given by \(\alpha\cdot\ket{0} + \beta\cdot\ket{1}\) with \mbox{complex-valued} amplitudes \( \alpha, \beta \). Following this description, \( |\alpha|^2 \) and \( |\beta|^2 \) are the probabilities to measure the base state \ket0 or \ket1, respectively, and, therefore, their sum \( |\alpha|^2 + |\beta|^2 \) has to be equal to \(1\).

Manipulation of a quantum state is achieved by a sequence of ``simple'' quantum operations (also denoted quantum gates) that are described by \emph{unitary matrices} acting on one or more qubits each. Such sequences of quantum operations are represented by circuit diagrams that indicate (from left to right) which operations are applied to which qubits.

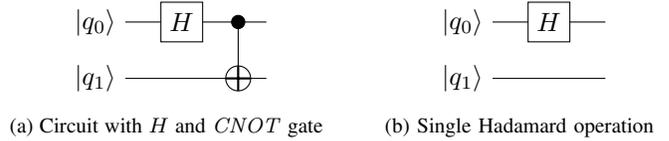
\begin{figure}[t]
	\centering
	\begin{subfigure}[t]{0.47\linewidth}
		\centering
		\begin{tikzpicture}[gate/.style={draw,fill=white,minimum size=1.5em},
		control/.style={draw,fill,shape=circle,minimum size=5pt,inner sep=0pt},
		cross/.style={path picture={\draw[thick,black](path picture bounding box.north) -- (path picture bounding box.south) (path picture bounding box.west) -- (path picture bounding box.east);}},
		target/.style={draw,circle,cross,minimum width=0.3 cm},
		x=0.75cm, y=0.75cm]
		\node[anchor=east] at (0,1) {$\ket{q_0}$};
		\node[anchor=east] at (0,0) {$\ket{q_1}$};
		
		\node[gate] at (1,1) {\(H\)};
		\node[control] at (2,1) {};
		\node[target] at (2,0) {};
		\draw (2,1) -- (2,0);
		
		\begin{scope}[on background layer]
		\draw (0,0) -- (2.5,0);
		\draw (0,1) -- (2.5,1);
		\end{scope}
		\end{tikzpicture}
		\caption{Circuit with \(H\) and \(\mathit{CNOT}\) gate}
		\label{fig:quantum-circuit-tikz}
	\end{subfigure}
	\hfill
	\begin{subfigure}[t]{0.47\linewidth}
		\centering
		\begin{tikzpicture}[gate/.style={draw,fill=white,minimum size=1.5em},
		control/.style={draw,fill,shape=circle,minimum size=5pt,inner sep=0pt},
		cross/.style={path picture={\draw[thick,black](path picture bounding box.north) -- (path picture bounding box.south) (path picture bounding box.west) -- (path picture bounding box.east);}},
		target/.style={draw,circle,cross,minimum width=0.3 cm},
		x=0.75cm, y=0.75cm]
		\node[anchor=east] at (0,1) {$\ket{q_0}$};
		\node[anchor=east] at (0,0) {$\ket{q_1}$};
		
		\node[gate] at (1,1) {\(H\)};
		
		\begin{scope}[on background layer]
		\draw (0,0) -- (2,0);
		\draw (0,1) -- (2,1);
		\end{scope}
		\end{tikzpicture}
		\caption{Single Hadamard operation}
		\label{fig:single-qubit}
	\end{subfigure}
	\caption{Quantum circuits}
	\vspace*{-5mm}
\end{figure}

\begin{example}
	Common operations performed on single qubits are the NOT operation X, the \emph{Hadamard} operation H to set a qubit into superposition, and the \emph{phase shift} operation T. The corresponding unitary matrices are defined as
	\[ 
		X = \begin{bmatrix}0 & 1 \\ 1 & 0\end{bmatrix}, \quad
		H = \frac{1}{\sqrt{2}} \begin{bmatrix*}[r]1 & 1 \\ 1 & -1\end{bmatrix*} \text{, and }\quad
		T = \begin{bmatrix}1 & 0 \\ 0 & e^\frac{i\pi}{4}\end{bmatrix}. \\
	\]
	In addition to the single qubit operations, there are also controlled operations such as the controlled NOT (CNOT) operation shown in the right-hand side of the quantum circuits depicted in Figure~\ref{fig:quantum-circuit-tikz} (next to the Hadamard operation). Here, the qubit \(q_1\) will only be negated iff \(q_0\) is in the \ket1 state. Due to the superposition introduced by the Hadamard operation, both qubits become \emph{entangled}, i.e.,~a measurement of one qubit affects the quantum state of the other qubit as well.
\end{example}

The overall functionality of a quantum circuit is determined by successively multiplying all gate matrices of the circuit. Hence, all gate matrices must have a dimension of $2^n \times 2^n$ (assuming an $n$-qubit system). If a gate matrix operates on a subset of the qubits only, the $2\times 2$ identity matrix $I_2$ is assumed for the other qubits. Forming the Kronecker product of these matrices results again in a matrix of size $2^n\times 2^n$ (assuming an $n$-qubit quantum system).

\subsection{Decision Diagrams for Quantum Computing}
\label{sec:qmdd}

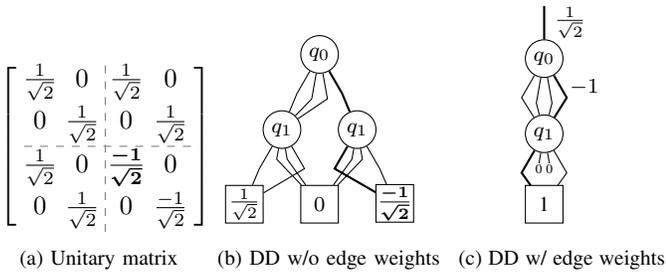
\begin{figure}
	\begin{subfigure}[t]{0.3\linewidth}
		\centering
		\begin{tikzpicture}
			\matrix (mymatrix) [matrix of math nodes, inner sep=1pt, left delimiter={[},right delimiter={]},nodes={outer sep=0pt, inner sep=0pt}, column sep=3pt, row sep=2pt]
			{
				\frac{1}{\sqrt{2}} &                  0 &  \frac{1}{\sqrt{2}} &                   0 \\
				0 & \frac{1}{\sqrt{2}} &                   0 &  \frac{1}{\sqrt{2}} \\
				\frac{1}{\sqrt{2}} &                  0 & \boldsymbol{\frac{-1}{\sqrt{2}}} &      0 \\
				0 & \frac{1}{\sqrt{2}} &                   0 & \frac{-1}{\sqrt{2}} \\
			};
			\draw[gray, dashed] (mymatrix.south) -- (mymatrix.north);
			\draw[gray, dashed] (mymatrix.east) -- (mymatrix.west);
		\end{tikzpicture}
		\caption{Unitary matrix}
		\label{fig:unitary-matrix}
	\end{subfigure}
	\hfill
	\begin{subfigure}[t]{0.35\linewidth}
		\centering
		\begin{tikzpicture}[terminal/.style={draw,rectangle,inner sep=0pt,xshift=-.5cm},
		font={\footnotesize},
		vertex/.style={draw,circle,inner sep=0pt,minimum width=0.5cm,minimum height=0.5cm},
		zeroterm/.style={below,inner sep=0pt,font=\tiny}]
		\matrix[matrix of nodes,ampersand replacement=\&,every node/.style={vertex},column sep={1cm,between origins},row sep={1cm,between origins}] (qmdd) {
			\node [xshift=.5cm] (n1) {$q_0$}; \& \&\\
			\node (n2) {$q_1$};  \& \node (n2a) {$q_1$}; \&\\
			\node[terminal] (t1){$\frac{1}{\sqrt{2}}$}; \& 	\node[terminal] (t0){\phantom{$\frac{1}{\sqrt{2}}$}}; \& 	\node[terminal, outer sep=0pt] (tm1){$\boldsymbol{\frac{-1}{\sqrt{2}}}$};\\
		};
		
		\draw (n1) -- ++(240:0.6cm) -- (n2);
		\draw (n1) -- ++(260:0.6cm) -- (n2);
		\draw (n1) -- ++(280:0.6cm) -- (n2);
		\draw[thick] (n1) -- ++(300:0.6cm) -- (n2a);
		
		\draw (n2) -- ++(240:0.6cm) -- (t1);
		\draw (n2) -- ++(260:0.4cm) -- (t0);
		\draw (n2) -- ++(280:0.4cm) -- (t0);
		\draw (n2) -- ++(300:0.6cm) -- (t1);
		
		\draw (n2a)[thick] -- ++(240:0.6cm) -- (tm1);
		\draw (n2a) -- ++(260:0.4cm) -- (t0);
		\draw (n2a) -- ++(280:0.4cm) -- (t0);
		\draw (n2a) -- ++(300:0.6cm) -- (tm1);
		\node at (t0) {$0$};
		\end{tikzpicture}
		
		\caption{DD w/o edge weights}
		\label{fig:dd-wo-edge-weights}
	\end{subfigure}
	\hfill
	\begin{subfigure}[t]{0.31\linewidth}
		\centering
		\begin{tikzpicture}[terminal/.style={draw,rectangle,inner sep=0pt},
		font={\footnotesize},
		vertex/.style={draw,circle,inner sep=0pt,minimum width=0.5cm,minimum height=0.5cm},
		zeroterm/.style={below,inner sep=0pt,font=\tiny}]
		\matrix[matrix of nodes,ampersand replacement=\&,every node/.style={vertex},column sep={1cm,between origins},row sep={1cm,between origins}] (qmdd) {
			\node (n1) {$q_0$}; \\
			\node (n2) {$q_1$}; \\
			\node[terminal, outer sep=0pt] (t){1};\\
		};
		\draw (n1) -- ++(240:0.6cm) -- (n2);
		\draw (n1) -- ++(260:0.6cm) -- (n2);
		\draw (n1) -- ++(280:0.6cm) -- (n2);
		\draw[thick] (n1) -- ++(300:0.6cm) node[right,midway] {$-1$} -- (n2);
		
		\draw[thick] (n2) -- ++(240:0.6cm) -- (t);
		\draw (n2) -- ++(260:0.4cm) node[zeroterm]{0};
		\draw (n2) -- ++(280:0.4cm) node[zeroterm]{0};
		\draw (n2) -- ++(300:0.6cm) -- (t);
		\draw[thick] ($(n1)+(0,0.7cm)$) -- (n1) node[right, midway]{$\frac{1}{\sqrt{2}}$};
		\end{tikzpicture}
		\caption{DD w/ edge weights}
		\label{fig:qmdd}
	\end{subfigure}
	\caption{Representations for $U = H \otimes I_2$.}
	\label{fig:representations}
	\vspace*{-5mm}
\end{figure}

Since the unitary matrices representing quantum functionality grow exponentially with respect to the number of qubits, i.e.,~the size of the quantum system, representations based on two-dimensional arrays quickly become infeasible. 
However, the structure of many explicitly used matrix instances offers the potential to exploit redundancies and, by this, allow for a drastically more compact representation (while still preserving an efficient manipulation). As in the classical realm, this is exploited by decision diagrams such as \cite{WLTK:2008,abdollahi2006analysis,VRMH:2003,MT:2006,DBLP:journals/tcad/NiemannWMTD16}.
These decision diagrams represent the matrix as a directed acyclic graph, where identical sub-matrices are combined in a shared graph structure. 
Further potential for redundancies can be achieved by sharing (sub-)graphs which are structurally equivalent and only differ by a common factor (to be annotated as a weight to the corresponding edges). 
The following example illustrate the main ideas:

\begin{example}
	Figure~\ref{fig:representations} shows different representations of the quantum functionality shown in Figure~\ref{fig:single-qubit}, i.e.,~described by the Kronecker-product \({H \otimes I_2}\). In Figure~\ref{fig:unitary-matrix}, the corresponding unitary matrix is shown, while Figure~\ref{fig:dd-wo-edge-weights} shows a corresponding decision diagram structure. In this graph, the \emph{root node} labeled \(q_0\) represents the whole matrix, while the four outgoing edges point to nodes that represent the top-left, top-right, bottom-left, and bottom-right sub-matrices (from left to right, hinted in the matrix by dashed lines). These $2 \times 2$ sub-matrices (representing functionality with respect to~$q_1$ only and, hence, represented by nodes labeled $q_1$) are further decomposed---yielding single complex values (or $1 \times 1$ sub-matrices) represented by terminal nodes.	
	As shown in Figure~\ref{fig:unitary-matrix}, the top-left, top-right, and bottom-left sub-matrices are identical and, therefore, are represented by the same node in Figure~\ref{fig:dd-wo-edge-weights} (the leftmost node labeled \(q_1\)). 
	
	Furthermore, the bottom-right sub-matrix is structurally equivalent to the other sub-matrices and only differs by a common factor, namely~$-1$. Hence, the matrix can be represented even more compactly as shown in Figure~\ref{fig:qmdd}. Here, additional edge weights are employed while the eventual matrix entry is determined by multiplying all edge weights from the root node to the terminal.\footnote{For the sake of readability, edge weights~\(1\) are omitted and weights~\(0\) are depicted as stubs.} This does not only allow to share the node representing the bottom-right sub-matrix, but also to boil down the number of required terminals to one (the actual complex values are now determined as product of the edge weights). As an example, the matrix element highlighted bold in Figure~\ref{fig:unitary-matrix} is determined by multiplying all weight on the bolded path in Figure~\ref{fig:qmdd}. Eventually, this yields a substantially more compact representation as originally given by the matrix.
\end{example}

The compaction can be further improved by \emph{normalizing} the nodes of the decision diagram (such as described in~\cite{DBLP:conf/rc/NiemannWD13}). This is commonly achieved by using the leftmost non-zero weight of the outgoing edges as a normalization factor.
All outgoing edges of the corresponding node are then divided by this factor, which is propagated towards the incoming edges to ensure products on all paths do not change.
This normalization scheme further provides a canonical, i.e.,~unique, representation of unitary matrices (as proven in~\cite{DBLP:journals/tcad/NiemannWMTD16}).

\section{Implementing\\Decision Diagrams for Quantum Computing}
\label{sec:implementation}

In this section, we discuss how to implement DD packages for the quantum domain in an efficient fashion. To this end, we utilize concepts such as~\emph{unique tables}, dedicated \emph{garbage collection}, or \emph{compute tables} which are already taken for granted in DD packages for the classical domain~\cite{somenzi2015cuddpackage,herbstritt2004wld,knuth2009art}. 
In addition to that, we show that just adopting those concepts is not sufficient for a highly efficient quantum DD package and additional implementation techniques are required which allow for efficiently handling complex numbers and sub-factors thereof (essential for the edge weights discussed in the previous section).

\subsection{Established Implementation Techniques}
The key concepts required for implementing DD packages for the classical domain in an efficient fashion have been introduced in the~'90s~\cite{Bry:86,BRB:90,SB:96} and leveraged the development of efficient packages such as \cite{somenzi2015cuddpackage,herbstritt2004wld,knuth2009art} (which served as basis for efficient methods for verification or synthesis). Many of these concepts can be directly re-used decision diagrams in the quantum domain, namely:

\begin{itemize}
	\item \emph{Unique tables} which store the nodes of the decision diagram and allow to efficiently detect redundancies in the structure. The unique table is realized as two-dimensional hash table (one for each variable in the decision diagram), where each bucket of the table contains a linked list of DD nodes with an equal associated hash value for each variable.
	Before any new DD node is inserted, it is checked whether the node already exists in the unique table.\footnote{The hash function should distribute the nodes equally to the buckets to reduce the number of collisions at lookup.}
	A DD node is returned by means of a pointer to the respectively allocated memory to allow dereferencing in constant time. Hence, it can be denoted as \emph{strong canonical form}~\cite{BRB:90}, i.e.,~a unique representation.

	\item Dedicated \emph{garbage collection} which frequently removes unused DD nodes from the unique table and, by this, allows for a fast insertion of new nodes as well as for keeping the memory usage low. Here, \emph{reference counting} is used to keep track which node shall be kept in the unique table. In order to avoid recurring memory allocations and deallocations, free DD nodes (i.e.,~allocated nodes that are currently not stored in the unique table) are stored in a list and corresponding memory is allocated en block.\footnote{Whenever the list is empty, we allocate enough memory to instantiate several new DD nodes.} Nodes that are removed from the unique table are appended to this list. To reduce the overhead caused by the lists (the list of free nodes as well as the lists in the buckets of the unique table), DD nodes themselves contain a pointer to the next element in the list.
	
	\item \emph{Compute tables} which cache the results of operations that are repeatedly conducted on the same DD nodes. Since decision diagrams exploit redundancies by using shared nodes, this proves very beneficial as many operations are frequently repeated and, hence, do not have to be recomputed. Compute tables are realized as hash tables, where the DD nodes denoting the operands (and possibly the type of operation) are used to determine the hash key~\cite{BRB:90}. The entries in the hash table might contain a single result that may be overwritten, or linked lists to store all computations with the same hash key to increase the \emph{hit rate} (e.g.,~with an LRU strategy as proposed in~\cite{SB:96}). Note that the compute tables have to be (partially) invalidated/cleared after garbage collection, since operands (referenced by a pointer) may be overwritten after the respective nodes have been removed from the unique table.

\end{itemize}

In fact, all these concepts can be (and have been) directly incorporated into implementations of decision diagrams for quantum computing reviewed in Section~\ref{sec:qmdd}. However, DD packages for quantum computing additionally have to handle the frequently occurring complex numbers. This constitutes \emph{the} major obstacle towards an efficient implementation of a \mbox{fully-fletched} DD package for the quantum domain as discussed in the next section. 

\subsection{Handling Complex Numbers}
\label{sec:complex_numbers}

As discussed in Section~\ref{sec:background}, weights that are attached to DD nodes offer the possibility for further compaction. 
In the classical domain, having such edge weights does not constitute an issue from the implementation perspective, since they are (tuples of) integers~\cite{bryant1995verification,herbstritt2004wld,Bry:85,Min:93,somenzi2015cuddpackage,knuth2009art}---again, a strong canonical form (a unique representation). This allows to use the efficient concepts outlined above, since computing unambiguous hash keys for DD nodes (containing weights attached to outgoing edges) is still possible.

In the quantum domain, however, weights are formed out of complex numbers. From a mathematical perspective, this does not causes problems since complex numbers also provide a strong canonical form.
However, it introduces severe challenges from an implementation perspective where machine accuracy is limited and, hence, complex numbers are approximated---yielding to numerical errors in computations. In fact, changing one complex number attached as weight to an outgoing edge by a tiny fraction (e.g.,~by flipping the least significant bit of the mantissa of the real or imaginary part) may yield a completely differently computed hash key. Accordingly, redundancies might remain undetected by the \mbox{DD package} since the node is searched in the wrong bucket of the unique table---causing a substantially  larger decision diagrams even though redundancies are actually present.
Additionally considering that weights represent sub-factors of complex numbers further increase the possibilities of corresponding numerical instabilities.\footnote{A more detailed discussion of this issue is provided in~\cite{zulehner2019accuracy}.}

To overcome this issue, one can represent complex numbers as two quadratic irrational numbers of the form \mbox{$\frac{a+b\sqrt{2}}{c}$ with $a,b,c \in \mathbb{Z}$} for the real and imaginary part~\cite{MTG:2006,DBLP:books/daglib/0027785}---severely limiting the number of possible complex numbers and, thus, not allowing for representing arbitrary quantum functionality.
Alternatively, one can use an additional table to store complex numbers, where an edge weight is then represented by an index of this table holding the corresponding number~\cite{DBLP:books/daglib/0027785,niemann2017efficient}. The table is then maintained in a fashion that numbers that do not differ by more than a tolerance value~$\epsilon$ share an entry in this table.\footnote{Note that, in~\cite{DBLP:books/daglib/0027785}, this $\epsilon$ is a relative tolerance value, while it is absolute in~\cite{niemann2017efficient}.}

However, while using a lookup table for complex numbers indeed allows for representing arbitrary quantum functionality, it introduces several new problems when aiming for an efficient implementation.
Thus, rather straightforward implementations of the respective DD concepts are available only. In the remainder of this paper, we identify arising difficulties and propose new implementation techniques which, for the first time, explicitly allow for implementing complex-valued edge weights in an efficient fashion. 

\section{Efficient Handling of Complex Edge Weights}
\label{sec:handling-complex-weights}

As discussed above, handling complex-valued edge weights (and complex numbers in general) in an efficient fashion is a key to a fully-fletched DD package for quantum computing. To this end, we propose such techniques in this section---covering arising issues like numerical instabilities caused by $\epsilon$, how to realize an efficient lookup of complex numbers that considers~$\epsilon$ (thus, providing a strong canonical form), as well as how operations on DDs can be handled efficiently.

\subsection{Obtaining Numerical Stability}

Design tasks for quantum computing (like synthesis, verification, or simulation) heavily rely on multiplying unitary matrices either with each other or with vectors. From a numerical perspective, these operations are not critical, since the multiplication with a unitary matrix is a well conditioned operation---even when the individual entries of the matrix are determined as product of several factors (as done in an edge-valued decision diagram). However, this changes when introducing the tolerance value $\epsilon$ as discussed above to detect redundancies. More precisely, some factors of an entry in the unitary matrix might be significantly rounded.

This problem becomes evident when using a normalization scheme (to gain canonicity of DD nodes) as described in~\cite{DBLP:journals/mvl/MillerFT07}. Here, all weights of outgoing edges are simply divided by the leftmost non-zero outgoing edge weight while propagating this extracted factor to the parent nodes (cf.~Section~\ref{sec:background})---edge weights in the decision diagrams are likely to become either rather large or rather small. 
If the real and the imaginary part of an edge weight are now close to~\(0\) (in the interval~\([-\epsilon,\epsilon]\)), the weight is rounded to~0. By this, a sub-tree of the decision diagram is possibly pruned by setting several entries in the matrix to zero---ending up with a huge round-off error and numerical instabilities.

A much better numerical stability is reached by changing the utilized normalization scheme. We propose to divide all weights of the outgoing edges of a DD node by the weight with the largest absolute value. If several outgoing edges have attached weights with equal absolute values (as can be seen in Figure~\ref{fig:qmdd} for the node labeled~\(q_0\)), we divide by the leftmost of these weights to preserve canonicity. With this normalization scheme, it is guaranteed that all edge weights in the decision diagram have an absolute value between~0 and~1 (and, thus, also the absolute values of the real and imaginary parts are between~0 and~1)---making it less likely that a factor unintendedly rounds to~0. While this indeed helps to increase numerical stability, one can additionally exploit the knowledge that all occurring complex numbers are either on or inside the unit circle to store them efficiently. This is discussed in the next section.

\subsection{Looking up Complex Numbers}

As discussed in Section~\ref{sec:complex_numbers}, hashing complex numbers is usually not possible for many quantum functions due to rounding errors caused by the limited machine accuracy. Hence, different methods are required that allow for a unique and efficient lookup of complex numbers while still considering the tolerance value $\epsilon$. 

The general idea for an efficient lookup is to exploit the fact that numbers can be sorted. Since there does not exist a total order for complex numbers, we split them into their real and imaginary parts. These real-valued parts are then stored separately in a lookup-table.\footnote{Note that a separate insertion of the real and imaginary part may also reduce the overall numbers to be stored since the same number might occur as real/imaginary part in several complex numbers.} A complex number is then represented by a pair of pointers to elements in the lookup-table---a strong canonical form. Moreover, we store only the absolute value of the real and imaginary part and \enquote{hide} the sign bit in the pointer to the respective entry of the table. This additionally allows to conduct operations like multiplication with constants like $-1$, $i$, or $-i$, as well as computing the complex conjugate just by flipping bits and/or swapping pointers.

\vspace{200cm}

To realize a lookup table for real-valued entries, we exploit the knowledge of the normalization scheme discussed above. In fact, it is guaranteed that all numbers of the table are within the interval~$[0,1]$.\footnote{Note that this also holds for the real and imaginary part of the complex weight attached to the edge pointing to the root node, since we deal with unitary matrices.} To allow for an efficient lookup, we split this interval into $N$ equally distributed chunks.\footnote{Note that other strategies for splitting the interval~$[0,1]$ are possible.} These chunks are represented by entries in an array of size \(N\), where each entry initially contains an empty list for occurring numbers in the respective interval (similar to the buckets in a hash table). When a new real number $r$ shall be inserted, the corresponding bucket is traversed. If one of the numbers in this list is equal to $r$ (considering the tolerance value $\epsilon$), a pointer to the respective number is returned. Otherwise, $r$ is inserted into the bucket. 

However, one has to be careful since, by allowing a tolerance value of $\epsilon$, a sufficiently close number might be located in one of the neighboring buckets. More precisely, if $r-\epsilon$ or $r+\epsilon$ exceeds the border of the considered interval (represented by a bucket), one has to additionally look for a number sufficiently close to $r$ in the the corresponding neighboring bucket. For performance reasons, we propose to return the first value found which deviates less than $\epsilon$ from $r$. Alternatively, to improve numerical stability, one can continue searching for a value that is even closer to $r$.\footnote{Note that there exist at most two numbers in the lookup table that are closer than $\epsilon$ to $r$.}

We employ the implementation techniques used for the unique table (for storing DD nodes) also for the lookup table (for storing real numbers) to allow for an efficient realization. 
This includes reference counting to keep track which entries are still required to be in the lookup table, a list of ``free'' numbers, as well as a procedure to allocate memory for several new entries at once. Whenever garbage collection is conducted on the unique table, we additionally run a garbage collection routine on the lookup table to remove entries (appending them to the list of free entries) with a reference count of~0. Moreover, numbers can easily be  dereferenced, since pointers are returned.

By using a lookup table as discussed above, inserting a value $r$ has complexity $\mathcal{O}(1)$ if $N$ is chosen suitably and the (inserted) numbers are distributed equally in the interval~$[0,1]$ (avoiding long collision chains). However, like a hash table, a worst-case complexity of $\mathcal{O}(n)$ results when all \(n\) entries in the table are stored in the same bucket. 
As alternative to a table, one can also use a self-balancing binary tree like an AVL tree~\cite{sedgewick1983algorithms} or a red-black tree~\cite{guibas1978dichromatic} to store the real numbers. Then, the lookup of a number $r$ requires $\mathcal{O}(\log n)$, where $n$ is the number elements in the tree. This is again possible, since real numbers can be totally ordered, but additionally requires overhead for re-balancing the tree.
Eventually, a combination of both---a lookup table where all entries stored in one bucket are realized as a self-balancing binary tree---provides a compromise of both ideas from a complexity-theoretic consideration.

However, even with an efficient implementation of a lookup for complex (real) numbers, several other issues have to be considered when conducting operations on decision diagrams. These are discussed in the following section.

\subsection{Conducting Operations on Decision Diagrams}

In this section, we discuss how to efficiently handle complex-valued edge weights when conducting operations on decision diagrams. To this end, we first analyze arising issues (from using a lookup table for complex numbers) that might impair efficiency, namely:
\begin{itemize}
	\item Each intermediate computation on complex numbers requires to perform a lookup in the table and, additionally, may cause rounding. Such intermediate values occur before normalizing a DD node.
	
	\item Sub-results might contain complex numbers with absolute value larger than~1 (before normalizing a DD node).
	
\end{itemize}
The first issue may affect the efficiency since the number of entries in the lookup table grows significantly (by inserting intermediate values that are not used anymore afterwards). Moreover, intermediate computations are not conducted as accurately as possible (by performing a lookup that considers~$\epsilon$). The second issue even prohibits the use of a lookup table as introduced above, since it is not guaranteed that all entries are in within interval~$[0,1]$. However, both issues can be resolved by introducing a cache for complex numbers that are used for storing intermediate results. Entries are taken from this cache whenever intermediate results are computed, and fed back when normalizing a newly computed DD node (before looking it up in the unique table).

In general, the efficiency of operations on decision diagrams results from a recursive formulation. Repeatedly re-evaluating the same (sub-)operation is avoided by using a compute table. Since each recursive call returns an already normalized DD node, cached complex numbers are only required for the current recursion level. 
Moreover, also the recursion levels above the current one may hold some cached complex numbers representing sub-results from other recursive calls (at most one for each other outgoing edge). From this, we can infer that a cache size linear to the number of variables is sufficient. Even more, by fixing the maximum number of variables in the decision diagram beforehand, a cache with fixed size for complex numbers (which is allocated at initialization of the package) is sufficient.

We implement this cache as list of real numbers and thereby utilize the same data structure as for entries in the lookup table discussed above. This allows to use entries from the lookup table and the cache interchangeably when computing, e.g.,~the product of two complex numbers. Cached complex numbers are allocated by taking two real numbers from the front of the list representing the cache. Before feeding them back to the cache when normalizing the computed DD node, we insert them into the lookup table (if no suitable number has been inserted yet). This is necessary since the lookup in the unique table requires a strong canonical form for the weights attached to the outgoing edges to detect redundancies. Since we look up complex numbers as late as possible, all computations are conducted with maximal precision. Intermediate results are not inserted into the lookup table for complex numbers, which keeps the number of entries in this table low and avoids repeated rounding (caused by~$\epsilon$) during computations.

On the downside, having a cache for complex numbers affects the compute tables that store sub-results. 
In fact, depending on the considered operation, an operand's weight (i.e.,~a DD edge weight) is either stored in the cache or in the lookup table. 
For complex numbers stored in the cache, we again have the problem that they do not necessarily preserve a strong canonical form. Hence, a completely different hash key may result from two slightly different complex numbers. This is not as critical as in the unique table, since it only may decrease the hit rate of the compute table. However, our internal evaluation has shown that the hit rate is hardly affected when rounding the complex numbers before computing the hash key.

\section{Resulting DD Package}
\label{sec:results}

In this section, we describe the resulting DD package for quantum computing when utilizing the implementation techniques introduced above.
As representative of a particular decision diagram type, we chose QMDDs as introduced in~\cite{DBLP:journals/tcad/NiemannWMTD16}, since this type incorporates all state-of-the-art concepts of decision diagrams for quantum computing reviewed in Section~\ref{sec:qmdd}.

As basis for our implementation served the QMDD package provided at \url{http://informatik.uni-bremen.de/agra/eng/qmdd.php}, which already utilizes the implementation techniques common for DD packages in the classical domain such as unique and compute tables, as well as garbage collection with reference counters. However, complex numbers are handled in a straightforward and inefficient fashion, namely:
\begin{itemize}
	\item Complex numbers are stored within an array of fixed size.
	\item The array is linearly traversed at each lookup, i.e.,~inserting a new number requires traversing the complete array.
	\item Complex numbers are inserted into the array at each step (causing rounding of intermediate results which may affect numerical stability).
	\item The normalization scheme does not consider numerical implications caused by $\epsilon$.	
\end{itemize}
All these shortcomings make the original QMDD package a proof of concept implementation (handling only small instances efficiently) rather than a fully-fletched DD package for quantum computing.

The contributions of this paper, i.e.,~the implementation techniques for an efficient handling of complex values, address these shortcomings and, hence, have been implemented on top of this package in C.
The resulting DD package is now publicly available at \url{http://iic.jku.at/eda/research/quantum_dd}.

To demonstrate the improved efficiency, we took established quantum functionality 
and generated the corresponding decision diagrams with the improved package. More precisely, we considered the \emph{Quantum Fourier Transform} (QFT~\cite{ekert1996quantum}) and the functionality of quantum circuits proposed by Google for quantum supremacy experiments~\cite{boixo2016characterizing}.
Since building the decision diagrams representing the respective functionalities is conducted by successively multiplying the individual quantum operations (also represented in terms of decision diagrams), this serves as a
representative case study on the efficiency of the package (after all, design automation methods heavily rely on matrix multiplication). 

\begin{table}
	\caption{Efficiency of the proposed techniques}\vspace{-.2cm}
	\label{tab:results}
	\scriptsize
	\resizebox{\linewidth}{!}{\begin{tabular}{l|S[table-format=2.0]|S[table-format=3.0]|S[table-format=7.0]|S[table-format=6.0,table-text-alignment=right,table-number-alignment=right]||S[table-format=>4.2,group-minimum-digits=4]||S[table-format=3.2]}
	name & $q$ & $\#\mathit{op}$ & $\mathit{size}$ & $\mathit{\#complex}$ &$t_{\mathit{original}}$ & $t_{\mathit{proposed}}$ \\\hline  
	\begin{filecontents*}{results.csv}
name,qubits,gates,time_old,nodes_end_old,complexnumbers_old,time_new,nodes_end_new,complexnumbers_new
Supremacy,16,50,0.00,72,238,0.00,80,33
Supremacy,16,70,4.04,6199,18803,0.06,6278,1486
Supremacy,16,75,1147.14,36107,225560,0.57,36161,6099
Supremacy,16,80,>3600.00,\textendash,\textendash,65.18,1195979,12670
Supremacy,20,70,0.06,455,3534,0.01,455,237
Supremacy,20,80,0.26,899,8010,0.02,899,546
Supremacy,20,89,1453.63,58251,97942,2.52,71105,6769
Supremacy,20,95,>3600.00,\textendash,\textendash,912.11,1742795,171586
Supremacy,25,100,0.15,912,6151,0.02,912,345
Supremacy,25,110,0.46,1751,12851,0.03,1751,1628
Supremacy,25,119,1282.28,58939,96647,2.63,58939,7380
Supremacy,25,120,>3600.00,\textendash,\textendash,3.51,365643,14878
QFT,15,120,3.98,32767,32785,0.39,32767,8195
QFT,16,136,17.48,65535,65554,0.86,65535,16386
QFT,17,153,67.02,131071,131091,1.64,131071,32771
QFT,18,171,295.48,262143,262164,3.50,262143,65539
QFT,19,190,1269.35,524287,524309,8.38,524287,131075
QFT,20,210,>3600.00,\textendash,\textendash,22.96,1048575,262147
\end{filecontents*}
\csvreader[late after line=\\, late after last line=\\,]{results.csv}
	{1=\name,2=\qubits, 3=\gates, 4=\tOld, 5=\nodesOld, 6=\complexOld, 7=\tNew, 8=\nodesNew, 9=\complexNew}
	{\name & \qubits & \gates & \nodesNew & \complexOld & \tOld & \tNew}
	\end{tabular}}

	~\\\raggedright
	$q$: number of qubits \qquad  \emph{\#op}: number of operations\\
	\emph{size}: size of the QMDD \qquad \emph{\#complex}: number of occurring complex values\\
	$t_{\mathit{original}}$: time when using the original QMDD package~\cite{DBLP:journals/tcad/NiemannWMTD16}\\
	$t_{\mathit{proposed}}$: time when incorporating the techniques proposed in this paper
	\vspace*{-5mm}
\end{table}

The obtained results are provided in Table~\ref{tab:results} which lists the number of qubits~$q$, the number of quantum operations~$\#\mathit{op}$, 
the \(\mathit{size}\) (i.e.,~the number of DD nodes) of the resulting decision diagram,
as well as the number of occurring $\mathit{\#complex}$ values throughout the computation.\footnote{Note that the number of occurring complex values can only be determined for the original package since all of them remain in an array during the computation.} 
Moreover, we list the runtime for building the decision diagram representing the respective functionality
when using the originally available implementation (denoted $t_{\mathit{original}}$) as well as when using the improved DD package utilizing the implementation techniques proposed in this work (denoted $t_{\mathit{proposed}}$).\footnote{Note that we had to slightly adjust the original package to allow storing more than 10\,000 different complex values. This, however, did not affect the runtime performance of the package.}

The obtained results clearly show that the proposed techniques allow to handle complex numbers much more efficiently than the original implementation. For example, consider the functionality of quantum circuits proposed for quantum supremacy experiments (proposed by researchers from Google~\cite{boixo2016characterizing}) with 16 qubits. Building the decision diagram representing the quantum functionality given by the first 50 operations can be handled by both packages in a fraction of a second. Considering the first 70 operations already results in a decision diagram with more than 6\,000 nodes and requires to deal with close to 20\,000 complex numbers. Even though these are not large numbers for decision diagrams, the original package already requires several seconds to build the decision diagram. When considering the first 75 operations, dealing with more than 36\,000 DD nodes and with approximately 225\,000 complex values requires almost 20 minutes---a task that can be solved in less than a second by the DD package resulting from this work. In fact, the limiting factor is not only the number of nodes, but also the number of complex values to deal with. Hence, the proposed implementation allows to build up the functionality of the first 80 operations---a rather large decision diagram with almost 1.2 million nodes---in a bit more than one minute, whereas the original implementation fails to do that within an hour. 

Overall, improvements in the performance of several orders of magnitude can be observed.
Hence, the evaluation demonstrates that complex values (especially when used as edge weights) can be handled much more efficiently by using the implementation techniques described in this paper. It is assumed that this performance improvement can directly be utilized in design automation methods relying on decision diagrams (such as those presented in~\cite{niemann2016logic,zulehner2017one,niemann2018cliffordt,viamontes2004high,zulehner2017advanced,zulehner2019matrix,WLTK:2008,yamashita2008ddmf,niemann2014equivalence}) just by replacing the currently used DD packages with the package proposed here (which is publicly available at \url{http://iic.jku.at/eda/research/quantum_dd}).

\section{Conclusions}
\label{sec:conclusions}

In this paper, we proposed implementation techniques for efficiently handling complex numbers in decision diagrams for quantum computing---especially when occurring as edge weights of the respective decision diagram.
By this, we leverage---in joint consideration of implementation techniques for decision diagrams in the classical domain developed decades ago---the development of a fully-fletched DD package for the quantum domain (publicly available at \url{http://iic.jku.at/eda/research/quantum_dd}).
The experimental evaluation showed that this package is indeed capable of handling complex numbers much more efficiently and allows constructing decision diagrams for established quantum functionality in significantly less runtime (up to several orders of magnitude). This performance boost is expected to be can passed to design automation methods utilizing decision diagrams just by incorporating this new package.

\section*{Acknowledgments}

This work has partially been supported by the LIT Secure and Correct System Lab funded by the State of Upper Austria.

\bibliographystyle{IEEEtran}
\bibliography{literature} 

\end{document}